\documentclass[aps,prd,nofootinbib,twocolumn,floatfix,unsortedaddress]{revtex4-1}

\usepackage{graphicx}
\usepackage{dcolumn}
\usepackage{subfigure}
\usepackage{times,mathptm}
\usepackage{float}
\usepackage{color}
\usepackage{amsmath,amsfonts}
\usepackage{mathptmx}
\usepackage{mathrsfs}
\usepackage{bbm}
\usepackage{bm}
\usepackage{xfrac}

\newcommand{\beq}{\begin{eqnarray}}
\newcommand{\eeq}{\end{eqnarray}}

\renewcommand{\b}{\beta}

\newcommand{\m}{\mu}

\newcommand{\s}{\sigma}

\newcommand{\dg}{\dagger}
\newcommand{\non}{\nonumber}

\newcommand{\rf}[1]{(\ref{#1})}
\newcommand{\ra}{\rightarrow}

\usepackage{ulem}

\begin{document}
\bibliographystyle{h-physrev5}

\title{Scaling properties of Wilson loops pierced by P-vortices} 

\author{Patrick Dunn}
\affiliation{Physics and Astronomy Dept., San Francisco State
University, San Francisco, CA~94132, USA}
\author{Jeff Greensite}
\affiliation{Niels Bohr International Academy, Blegdamsvej 17, DK-2100
Copenhagen \O, Denmark}
\altaffiliation[Permanent address: ]{Physics and Astronomy Dept., San Francisco State
University, San Francisco, CA~94132, USA}
\date{\today}
\begin{abstract}
   P-vortices, in an SU($N$) lattice gauge theory, are excitations on the center-projected $Z_N$ lattice.
We study the ratio of expectation values of SU(2) Wilson loops, on the unprojected lattice, linked to a single P-vortex,
to that of Wilson loops which are not linked to any P-vortices.  When these ratios are plotted versus loop area in
physical units, for a range of lattice couplings, it is found that the points fall approximately on a single curve, consistent with scaling.  We also find that the ratios are rather insensitive to the point where the minimal area of the loop is pierced by the P-vortex.

\end{abstract}

\maketitle

\section{\label{sec:intro}Introduction}

    The center vortex theory of confinement 
 \cite{'tHooft:1977hy,Mack:1978kr,Cornwall:1979hz,Feynman:1981ss,Nielsen:1979xu,Ambjorn:1980ms} is motivated by the fact that the asymptotic string tension associated with Wilson loops in group representation $r$, in a pure SU($N$) gauge theory, depends only on the $N$-ality of that representation, i.e.\ on the transformation properties of the Wilson loop holonomy with respect to the $Z_N$ center subgroup.   This behavior can be understood in ``particle'' language; e.g.\ the string which forms between a quark and antiquark in the adjoint representation of the gauge group is eventually broken by pair production of gluons, each of which binds to one of the quarks, resulting in two color singlet states consisting of a quark or antiquark bound to a gluon.   This explains why a zero N-ality loop (such as a Wilson loop in the adjoint representation) will have a vanishing asymptotic string tension. 
On the other hand, there should also be an explanation purely in ``field'' language, i.e.\ the dependence on $N$-ality ought to be explicable in terms of the field configurations which dominate the Euclidean functional integral at very large scales.    
Such field configurations must be organized in such a way that they generate string tensions for Wilson loops that
depend only on the $N$-ality of the loop.  To the authors' knowledge, center vortices are the only field configurations thus far proposed which have this property, and which do not have to appeal to some further color-screening mechanism in the particle picture.
     
     There is a great deal of lattice Monte Carlo evidence in favor of the center vortex mechanism that has accumulated over the years, cf.\ the reviews in refs.\ \cite{Greensite:2011zz,Greensite:2003bk}, which mainly cover the SU(2) case, and also the recent work in \cite{OMalley:2011aa} for the SU(3) gauge group. 
This data is based on the procedure of center projection in maximal center gauge.  One fixes to a gauge (direct or indirect maximal center gauge \cite{DelDebbio:1998uu}) which brings the link variables as close as possible, on average, to center elements.  In the direct maximal center gauge, the procedure is to maximize
\beq
   R = \sum_{x,\m} \mbox{Tr}[U_\m(x)]\mbox{Tr}[U^\dg_\m(x)]  
\eeq
via a relaxation technique which reaches a local maximum.  This can be regarded as fixing to Landau gauge in the adjoint representation.  Link variables $U_\m(x)$ are then projected to the a center element $z_\m(x) \in Z_N$ which is nearest to $U_\m(x)$ in the sense that 
\beq
         \lefteqn{\mbox{Tr}[(U_\m(x)-z_\m(x)\mathbbm{1}_N)(U^\dg_\m(x)-z^\dg_\m(x)\mathbbm{1}_N)]}
\non \\
      & & \qquad \qquad \qquad = 2N - \mbox{Tr}[z_\m(x) U_\m^\dg(x) + h.c]
\eeq
is minimized.  This mapping of configurations $U_\m(x) \ra z_\m(x)$ from the SU($N$) lattice to a $Z_N$ lattice is known as ``center projection."    String tensions computed on the center-projected lattice, in SU(2) lattice gauge theory, are known to have excellent scaling properties, and agree fairly well with the asymptotic string tensions computed on the unprojected lattice, a feature known as ``center dominance."\footnote{There are still some ambiguities, connected with Gribov copies, which can affect this result, c.f.\ \cite{Greensite:2003bk} for a discussion.}    The excitations on the projected $Z_N$ lattice are known as ``P-vortices."  We define a P-plaquette as a plaquette on the projected lattice whose value is an element of the $Z_N$ group different from unity.  P-vortices, in $D$-dimensions, are $D-2$ dimensional objects on the dual lattice, composed of elements (sites, links, or plaquettes in $D=2,3,4$ respectively) which are dual to P-plaquettes.  These are the center vortices of a $Z_N$ gauge theory.  The value of a Wilson loop on the projected lattice is simply unity times the product of P-plaquettes in the minimal area of the loop.

    The question is whether the location of P-plaquettes in the center-projected lattice is correlated to the value of
gauge-invariant Wilson loops on the unprojected lattice.  The evidence in favor of such a correlation is based on the
measurement of ``vortex-limited" Wilson loops.  A vortex-limited Wilson loop, $W_{n_1,n_2,...n_{N-1}}(C)$, is the expectation value of all loops on the unprojected lattice of shape $C$ whose minimal area contains, on the projected lattice, exactly $n_k$ P-plaquettes equal to center element $z_k = \exp[2\pi i k/N]$, for $k=1,...,N-1$.  In the center vortex picture, if we assume that the thick center vortices do not overlap the loop boundary, then the effect of the center vortices would be to contribute an overall phase factor
\beq
            z = \prod_{k=1}^{n-1} z_k^{n_k}
\eeq
to the value of the loop.  The area law falloff would be due to fluctuations in this phase factor, corresponding to fluctuations in the number of center vortices that are topologically linked to the loop.  In this article we
will only be concerned with SU(2) lattice gauge theory, where there is only one type of center vortex, and the vortex-limited Wilson loops are denoted $W_n(C)$.  The minimal areas of loops contributing to $W_n(C)$ are said to be
``pierced" by $n$ P-vortices.   We may also define $W_{odd (even)}(C)$ as the expectation value for Wilson loops pierced by an odd (even) number of P-vortices.  It was shown in the early work \cite{DelDebbio:1998uu} on this subject that $W_n$ depends very strongly on $n$, and the numerical evidence suggests that in the limit of large loop area
\beq
            {W_n(C) \over W_0(C)} \ra (-1)^n ~~~~ \mbox{and} ~~~~ {W_{odd}(C) \over W_{even}(C)} \ra -1 \ ,
\label{limits}
\eeq
This is consistent with the idea that a P-plaquette on the projected lattice is roughly correlated with the location of
a thick center vortex on the unprojected lattice, and that a thick center vortex, if topologically linked to loop $C$, will contribute a factor of $-1$ to the loop holonomy.  It should be noted that numerical simulations have also shown
 \cite{DelDebbio:1998uu} that vortex-limited Wilson loops $W_{0}(C),W_{even}(C)$ do not, by themselves, have an area-law falloff, and given that the ratios \rf{limits} are of $O(1)$, this lack of area law falloff must also hold true for $|W_1(C)|,W_2(C),|W_{odd}(C)|$.  For this reason, it is likely that this absence of area-law falloff holds true in general for all the vortex-limited Wilson loops. The standard Wilson loop expectation value is related to the vortex-limited loops via
\beq
     W(C) = \sum_n p_n(C) W_n(C) \ ,
\eeq
where $p_n(C)$ is the probability that the minimal area of a given planar loop $C$ contains $n$ P-plaquettes on the projected lattice.  If the 
$|W_n(C)|$ all have a perimeter-law falloff, as the numerical evidence suggests, then the area law can only be obtained from cancellations due to the sign differences among the different $W_n$'s, in complete accordance with the center vortex picture.  

      In the early SU(2) work,  the ratio $W_n(C)/W_0(C)$ was only computed at
a lattice coupling of $\b=2.3$.  There was no effort to test scaling, i.e.\ to check whether the Wilson loop ratios plotted versus area in physical units fall on a universal curve.  In a more recent study, Langfeld \cite{Langfeld:2003ev}  investigated the phase of vortex-limited Wilson loops in SU(3) lattice gauge theory at two lattice couplings, $\b=5.6, 5.8$, and obtained results which were roughly consistent with scaling.  In this article we will continue to work with SU(2) loop
ratios, as in the early work, but display a larger data set of loop areas at six different lattice couplings, which may give an improved sense of the scaling properties.

\section{$W_1/W_0$ scaling, and P-plaquette location}

\begin{figure}[t!]
\centerline{\scalebox{0.7}{\includegraphics{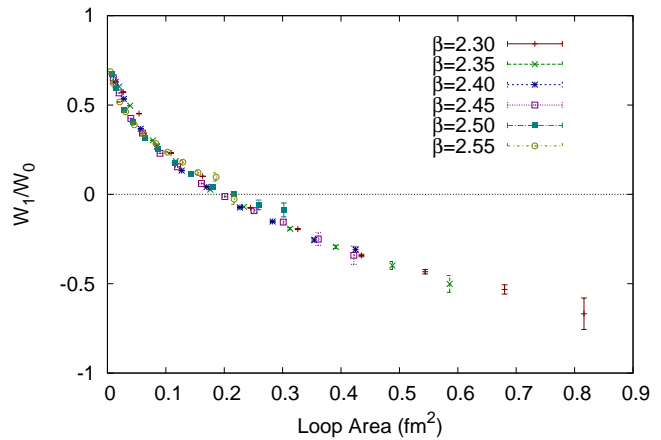}}}
\caption{Ratio of vortex-limited Wilson loop expectation values $W_1(C)/W_0(C)$ vs.\ loop area in physical units,
at various lattice couplings.}
\label{fig1}
\end{figure}

    Let $W_n[I,J]$ represent the expectation value of a vortex-limited loop, where the loop is a rectangular contour
of $I$ lattice units on one side, and $J$ lattice units on the other.  We will consider loops where $I=J$, and $|I-J|=1$.
The area of the loop in physical units is $IJa^2$, where the lattice spacing is given, as usual, by $a=\sqrt{\s_L/\s}$,
where $\s_L$ is the string tension in lattice units, and we take $\s= (440~\mbox{MeV})^2$.  Fig.\ \ref{fig1} displays our results
for $W_1[I,J]/W_0[I,J]$ vs.\ area $IJa^2$ in units of fm${}^2$, for lattice couplings ranging from $\b=2.3$ to
$\b=2.55$ on $24^4$ lattice volumes.  The data seems to fall roughly on the same curve, which indicates that the $W_1/W_0$ ratio is a physical observable of some kind.    The usual interpretation of these results is that a single P-plaquette found on the projected lattice, in the minimal area of loop $C$, indicates that loop $C$  is linked to a thick center vortex
on the unprojected lattice.  Of course, a small loop
on the unprojected lattice cannot be affected by the full center flux carried by a thick center vortex, so one does not
expect $W_1(C)/W_0(C)$ to equal $-1$ in that case.  This limit should be obtained, however, when the loop area grows much larger than the cross-sectional area of a vortex, so the single vortex linked to the loop does not overlap the boundary of the loop.  Fig.\ \ref{fig1} appears to be consistent with this expectation, although only for the lowest couplings are we able to measure loops which are large enough, in physical units, such that the $W_1/W_0$ ratio approaches the expected asymptotic value of $-1$.

\begin{figure}[thb]
\begin{center}
\subfigure[]  
{   
 \includegraphics[scale=0.25]{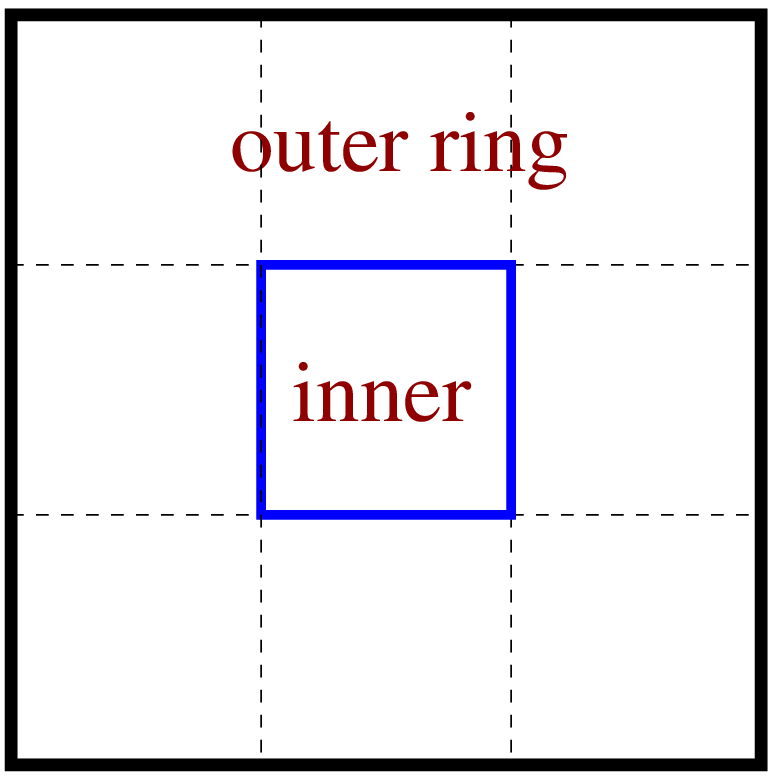}
}
\hspace{1.0cm}
\subfigure[]   
{  
 \includegraphics[scale=0.25]{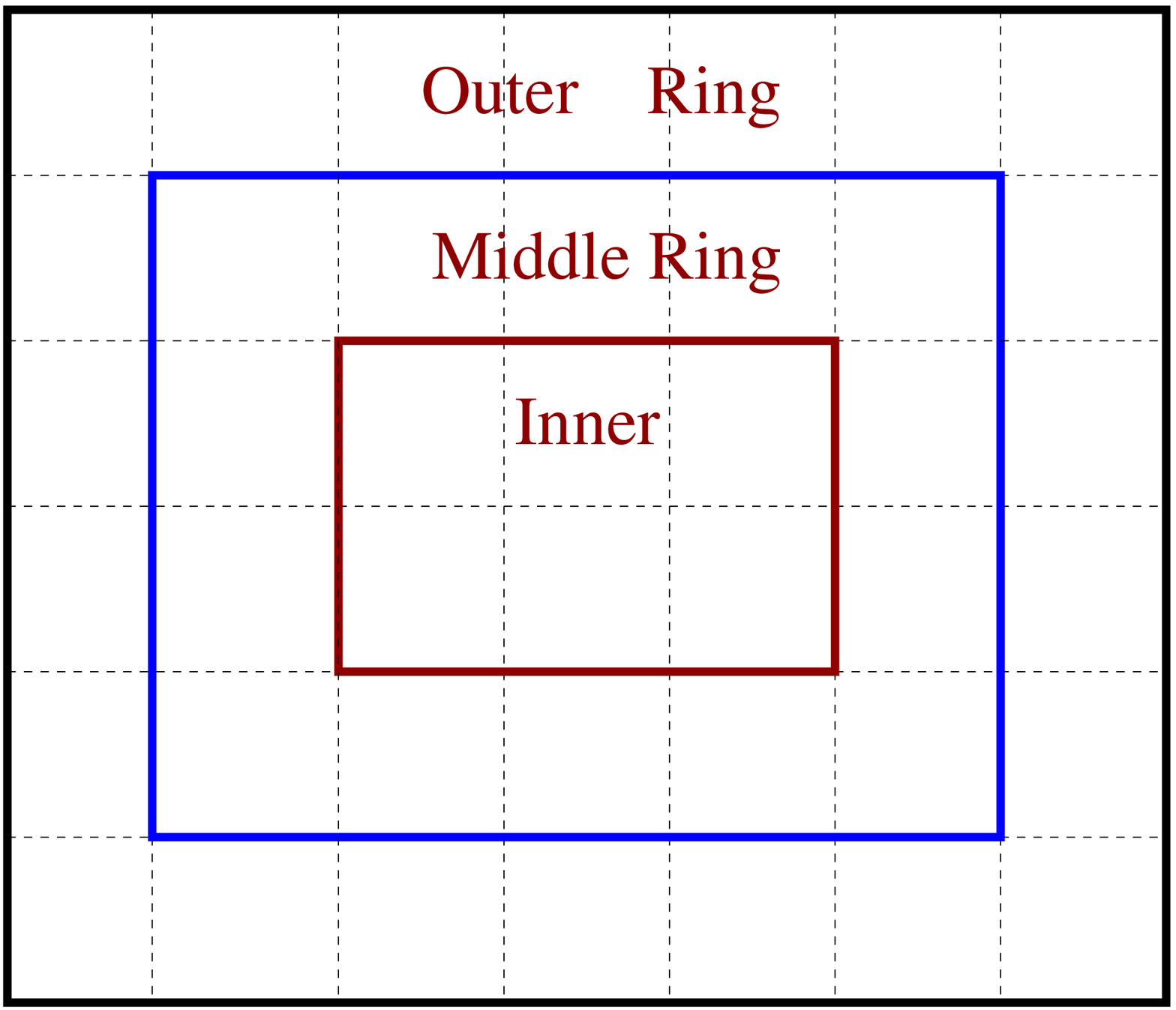}
}
\end{center}
\caption{Conventions for labeling P-plaquette positions.  Plaquettes in the plane of, but just outside the loop, bordering
the perimeter, are referred to as ``outside."}
\label{draw}
\end{figure}

    We next consider the question of how the ratio $W_1(C)/W_0(C)$ depends on the position of the P-plaquette
within the minimal area of loop $C$.  For this study we will also consider the case, which does not really belong to $W_1$, in which the P-plaquette lies in the plane of loop C but just outside the minimal area, bordering the perimeter of loop $C$.   The ratio of Wilson loops of this kind to $W_0$  will be represented by data points labeled ``outside."  Points labeled ``inside" are the usual ratios of $W_1/W_0$.  We then make the following distinctions:  Consider all plaquettes inside the minimal area of the loop, which border the perimeter.  These plaquettes belong to the minimal area bordered by $C$ and another rectangular loop $C_1$.  If the minimal area of $C_1$ is non-zero, then the data for $W_1/W_0$, with P-plaquettes in
this region between $C$ and $C_1$, is labeled ``outer ring."   Next, consider P-plaquettes in the minimal area of 
$C_1$ bordering the perimeter of $C_1$, and another rectangular loop $C_2$.  If the minimal area of $C_2$ is non-zero, these P-plaquettes belong to the "middle ring," and any P-plaquettes within the minimal area of $C_2$ are denoted
``inner."  If, on the other hand, the minimal area of $C_2$ is zero, then P-plaquettes in the minimal area of $C_1$ are themselves denoted ``inner."  Our conventions are illustrated in Fig.\ \ref{draw}.

\begin{figure}[htb]
\centerline{\scalebox{0.70}{\includegraphics{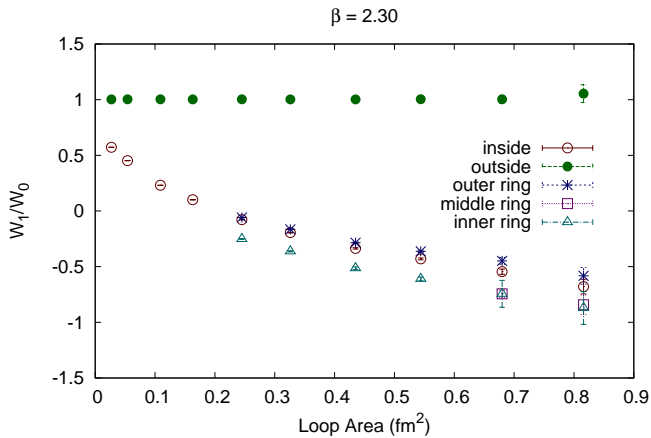}}}
\caption{Vortex-limited Wilson loop ratios vs.\ loop area in physical units, for specific positions of the P-vortex relative to the loop perimeter (see Fig.\ \ref{draw}), at $\b=2.3$.}
\label{fig2}
\end{figure}

\begin{figure}[htb]
\centerline{\scalebox{0.70}{\includegraphics{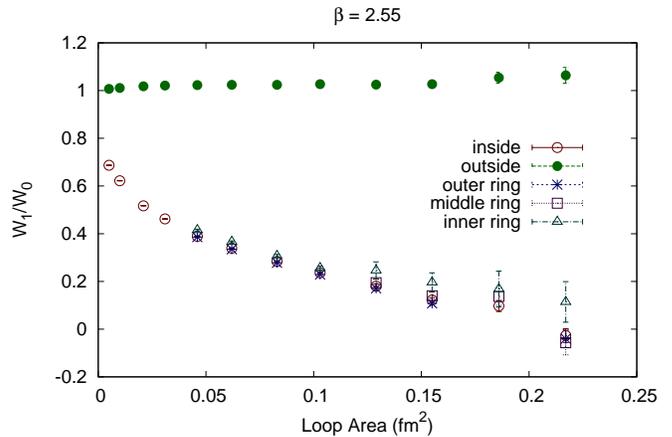}}}
\caption{Same as Fig.\ \ref{fig2}, at $\b=2.55$.}
\label{fig3}
\end{figure}

     The numerical results, shown for $\b=2.3$ in Fig.\ \ref{fig2} and $\b=2.55$ in Fig.\ \ref{fig3} are a little surprising, since
one would expect that if the location of a P-vortex were strongly correlated with the middle of a thick center vortex, then the ratio $W_1(C)/W_0(C)$, for a large loop, would systematically fall from outer to middle to inner.
While this does seem to be true at $\b=2.3$, it is not a very large effect, and is not observed at all at $\b=2.55$ (if anything, the expected order is reversed).   The really striking effect
is the dramatic dependence on whether a P-plaquette lies just outside, or just inside, the perimeter of the loop.
If the P-plaquette lies just outside the loop (data points labeled ``outside"), then the expectation value of the loop differs hardly at all from $W_0$.  In sharp contrast, if the P-plaquette lies anywhere within the minimal area of the loop, including at the loop perimeter, then the loop expectation value differs greatly from $W_0$.   This difference between loops with
one inside P-plaquette at the perimeter, and loops with one outside P-plaquette at the perimeter, increases with loop area, and it is quite remarkable in view of the insensitivity of $W_1/W_0$ to the location of the P-plaquette within the minimal area.

\section{Discussion}

   We have seen that the $W_1(C)/W_0(C)$ ratio scales reasonably well with $\b$, and also that there is an extremely
strong correlation between the expectation value of a Wilson loop, and whether a P-plaquette, bordering the perimeter,
lies just inside or just outside the minimal area.  There are a priori reasons to expect the scaling property, since, e.g., center-projected string tensions scale rather nicely \cite{DelDebbio:1998uu}, but such a strong distinction between P-plaquettes lying just inside or just outside the loop is a little surprising, especially for large loops at large $\b$.  For a thick center vortex, one would expect that the amount of center flux penetrating the minimal area of the loop would not be very different if the middle of the vortex were located just inside, or just outside, the loop perimeter.  Yet the trend of our data indicates that expectation values of large
loops depend very strongly on whether or not a single P-plaquette is located inside the loop, but, if inside, the loop expectation value is rather insensitive to exactly {\it where} inside.  This result would make perfect sense if
P-vortices were very strongly correlated with the position of center vortices on the unprojected lattice, and if those center vortices were only one lattice spacing wide.  But if that were the case, then the ratio $W_1(C)/W_0(C)=-1$  should be
obtained for even the smallest loops, and not just as an asymptotic limit.   On the other hand, if center vortices are rather thick in lattice units,  and the location of P-plaquettes and center vortices is only weakly correlated,  one would not expect such a striking difference in our values labeled ``outside" and ``outer ring,"  corresponding to a P-plaquette just outside or just inside the loop perimeter.  

    So it appears that the location of a P-plaquette is not a very good guide to the precise position of a thick center vortex.  On the other hand, the presence of a single P-plaquette anywhere inside the loop tells us that the sign of a large SU(2) Wilson loop is, on average, negative.  It seems that a P-plaquette inside a loop is strongly indicative of center flux passing through the loop, but does not give us much information about how that flux is distributed inside the minimal area.

\acknowledgments{This work is supported in part  by
the U.S.\ Department of Energy under Grant No.\ DE-FG03-92ER40711.}

\bibliography{dunn}

\end{document}